\documentstyle[12pt]{article}
\begin{document}
\vphantom{0}
\vskip1.1truein
\renewcommand{\theequation}{\thesection.\arabic{equation}}
\newcommand{\eqal}[2]{\begin{eqnarray} #2 \end{eqnarray}}
\newcommand{\eqarr}[2]{\begin{array} #2 \end{array}}
\newcommand{\eqn}[2]{\begin{equation} #2 \end{equation}}
\newcommand{\newsec}[1]{\setcounter{equation}{0} \section{#1}}
\newcommand{\no}{\nonumber}
\newcommand{\tr}{{\rm tr~}}
\newcommand{\Tr}{{\rm Tr~}}
\newcommand{\half} {\frac{1}{2}}
\newcommand{\cint} {\oint}
\newcommand{\CP} {\wp}
\newcommand{\CL} {{\cal L}}
\def\yboxit#1#2{\vbox{\hrule height #1 \hbox{\vrule width #1
\vbox{#2}\vrule width #1 }\hrule height #1 }}
\def\fillbox#1{\hbox to #1{\vbox to #1{\vfil}\hfil}}
\newcommand{\ybox}{{\lower 1.3pt
\yboxit{0.4pt}{\fillbox{8pt}}\hskip-0.2pt}}
\makeatletter
\ifcase\@ptsize
 \font\tenmsy=msbm10
 \font\sevenmsy=msbm7
 \font\fivemsy=msbm5
\or
 \font\tenmsy=msbm10 scaled \magstephalf
 \font\sevenmsy=msbm8
 \font\fivemsy=msbm6
\or
 \font\tenmsy=msbm10 scaled \magstep1
 \font\sevenmsy=msbm8
 \font\fivemsy=msbm6
\fi
\newfam\msyfam
\textfont\msyfam=\tenmsy  \scriptfont\msyfam=\sevenmsy
  \scriptscriptfont\msyfam=\fivemsy
\def\Bbb{\ifmmode\let\next\Bbb@\else
 \def\next{\errmessage{Use \string\Bbb\space only in math mode}}\fi\next}
\def\Bbb@#1{{\Bbb@@{#1}}}
\def\Bbb@@#1{\fam\msyfam#1}
\newcommand{\IZ}{{\Bbb{Z}}}
\makeatother
\leftline{Conformal Field Theory Techniques}
\leftline{in Large $N$ Yang-Mills Theory
\footnote{To be published in the proceedings of the May 1993 Carg\`ese
workshop on Strings, Conformal Models and Topological Field Theories.}}
\vskip.65truein

\hbox{\obeylines\baselineskip12pt\parskip0pt\parindent0pt\hskip1.1truein
\vbox{Michael R. Douglas
\vskip.1truein
Dept. of Physics and Astronomy
Rutgers University
}}
\vskip .65truein
\noindent
Following some motivating comments on large $N$ two-dimensional Yang-Mills
theory,
we discuss techniques for large $N$ group representation theory, using
quantum mechanics on the group manifold $U(N)$, its equivalence to
a quasirelativistic two-dimensional free fermion theory, and bosonization.
As applications, we compute the free energy for two-dimensional Yang-Mills
theory on the torus to $O(1/N^2)$,
and an interesting approximation to the leading answer for the sphere.
We discuss the question of whether the free energy for the torus
has $R\rightarrow 1/R$ invariance.

\vskip.65truein

\newsec{Introduction}

The first part of this article is an introduction to what might be called
``large N representation theory,''
Lie group representation theory with the focus on the limit
$N\rightarrow\infty$ of $SU(N)$ and the other classical groups.
This has many applications in physics and mathematics, and good
mathematical
introductions exist, which tie it to its applications in group theory,
soliton
theory, combinatorics, and so forth.
(See \cite{Jimbo}, 5.4 for a treatment very much like the one here;
see also \cite{Segal})
Now although it might seem that this theory would be invaluable for
studying
the large $N$ limit of models with $SU(N)$ symmetry, and some examples in
the
physics literature are  in \cite{DMP+Jevicki+Stone},
it does not get as much use as one might expect.
Whether this is simply because the language is taking time to standardize,
or
because the physics really is too diverse to capture in one formalism, I
leave
for the reader to judge.
However a model which seems made to order as an application is large $N$
two-dimensional Yang-Mills theory (YM$_2$ in the following).

I would like to make some comments about the potential physical relevance
of
this model, which is certainly a long way from realistic four-dimensional
models.  The starting point is the old idea that QCD could be reformulated
as a
string theory, which for many reasons can only be a free string in the
large
$N$ limit.
\cite{Polchinski}
Although our understanding of such things is primitive, it seems clear
that a
``QCD string'' if it exists is a very different type of string than those
studied as theories of quantum gravity, and that a low dimensional
solvable
model might provide a context in which we could discover and understand
such a
different type of string theory.  This line of thought has led to a
revival of
the study of two-dimensional Yang-Mills theory and QCD with the objective
of
developing a string representation which would generalize to higher
dimensions.
 The earliest work in this direction concentrated on reproducing Wilson
loop
expectation values; even in two dimensions these are non-trivial for
self-intersecting loops and evidence was found that their structure could
be
described by a sum of surfaces each with weight
$\exp -{\rm area}$, modified by local factors associated with features on
the
surface such as branch points. \cite{kkone}
More recently the partition function on a closed Riemann surface has been
studied and a complete set of rules derived which reproduces it as a sum
over
surfaces, again with additional features.
\cite{Gross}
The idea is not restricted to two dimensions and a $D$-dimensional lattice
formulation exists.
\cite{Kostov}

Besides explicit string constructions of this sort, much can be learned
from
comparing precise results from a field theory and candidate equivalent
string
theories.  Of course if we can get exact results for a theory we do not
really
need a string formulation but even crude results in higher dimensions are
valuable to show whether a string formulation could work at all.
The most important issue for the constructions of \cite{Gross,Kostov} is
that
they are essentially strong coupling expansions.  They improve on Wilson's
original expansion by replacing the expansion parameter $1/g^2$ by $\exp
-g^2$
($g$ is  a dimensionless bare lattice coupling) but still face the
essential
problem of the strong coupling expansion for gauge theory -- the continuum
limit requires the limit of weak bare coupling.
As it turns out there is an interesting two-dimensional case which
illustrates
the problems, namely YM$_2$ on the sphere.

Assuming the usefulness of these string representations, the next step
would be
to take a world-sheet continuum limit and hope that this still reproduces
the
Yang-Mills continuum limit.
The results of \cite{Gross} are similar to topological field theory
results,
and \cite{Bershadsky,DR} pointed out that for torus target space there is
a
natural candidate for comparison, the topological sigma model with torus
target
space (possibly with coupling to gravity), and that this worked for the
torus
world-sheet.  One consequence of this on arbitrary genus world-sheet would
be
$R\rightarrow 1/R$ duality invariance, and we will look for this at genus
two
in the YM$_2$ results.

This relation to topological theory seems very special to $D=2$, and in
higher dimensions we expect a string with less trivial dynamics.  However
the
generic string theory seems to have a trivial world-sheet continuum limit;
this
is the famous ``$c>1$'' or ``branched polymer'' problem which has been
argued
to be inevitable in a theory defined as a sum over world-sheets with
positive
weights. \cite{Ambjorn}
A sensible QCD string must escape this problem, and the additional
world-sheet
features have to play an essential role in this, leading to the question
of
which of the many features are essential and which are irrelevant.  One
would
like to understand this before trying to reproduce the string theory with
a
continuum world-sheet action.
\def\foo{
Already in $D=2$ this is non-trivial but unfortunately this seems like a
special case; the ``no folding'' rules of \cite{Gross} do not generalize
to the
$D>2$ formalism of \cite{Kostov}.}
It seems to me that this can only be properly understood in $D>2$;
nevertheless
the $D=2$ results do suggest that some features are more important than
others.

\newsec{Quantum Mechanics and Group Representations}

The prototypical system we study is quantum mechanics on the group
manifold
$U(N)$.
This allows us to quickly classify representations and derive the Weyl
character formula.  Physically this system is already interesting, since
it
describes the (global) degrees of freedom of two-dimensional Yang-Mills
theory.
We go on to discuss calculations of the YM$_2$ free energy on a Riemann
surface.

The natural Hamiltonian is
\eqn\qmham{
\label{qmham}
H = \tr \left(U{\partial\over\partial U}\right)^2 \equiv \sum_a E^a E^a.}
Here $E^a = \tr t^a U d/dU$ generates left rotations of $U$ and represents
the
Lie algebra $u(N)$.  Thus acting on a wave function which could be any
matrix
element of an irreducible representation $R$, $\psi(U)=D^{(R)}_{i\bar
j}(U)$,
$H = C_2(R)$, the second Casimir (normalized so that $C_2(\ybox)=N$).
In fact it is the unique invariant and purely second order linear
differential
operator on the group manifold, the Laplacian.

To classify representations we should find their characters
$\chi_R(U) = \tr D^{(R)}(U)$.
These will be wave functions invariant under
$\psi(U)\rightarrow \psi(gUg^{-1})$,
so we should make the change of variables $U_{ij}=g_{ik} z_k g^{-1}_{kj}$.
(This is familiar from the quantum mechanics of a hermitian matrix
\cite{BIPZ}
and for the group manifold case is much older, going back to
Harish-Chandra.
\cite{HCandHT})

The invariant volume element in these variables is
\eqn\haar{\sqrt{h} = |\Delta(z)|^2 = \tilde\Delta(z)^2}
where $\Delta(z)=\prod_{i<j}(z_i-z_j)$ and
$\tilde\Delta(z)=\prod_{i<j}\sin{\theta_i-\theta_j\over 2}
=\Delta(z)/\prod_i z_i^{(N-1)/2}$.
The ``radial'' components of the metric are simply $h_{ij} = \delta_{ij}$.
Thus on wave functions independent of $g$
\eqn\qmhamtwo{H = -\sum_i {1\over\tilde\Delta^2}{d\over d\theta_i}
\tilde\Delta^2{d\over d\theta_i}.}
We can rewrite this as
\eqn\qmhamtwob{H = -\sum_i \left[ {1\over\tilde\Delta}{d^2\over
d\theta_i^2}
\tilde\Delta - {1\over\tilde\Delta}
\left({d^2 \tilde\Delta\over d\theta_i^2}\right)
\right] .}

For hermitian matrix quantum mechanics $\Delta$ was a Vandermonde and the
second term, thanks to a non-trivial identity, gave zero.  Here, after a
similar calculation, the second term is found to equal $-N(N^2-1)/12$.
\cite{Dowker}

Thus, after redefining the wave functions by
$\psi\rightarrow\tilde\Delta\psi$,
we have a theory of $N$ free fermions on the circle.  The boundary
conditions
are also determined by this redefinition; they become periodic
(antiperiodic,
respectively) for $N$ odd (even).
An orthonormal basis for wave functions is Slater determinants
\eqn\slate{\psi_{\vec n} = \det_{i,j} z_i^{n_j}}
with energy $E = \sum_i n_i^2 - N(N^2-1)/12$.
The ground state has fermions distributed symmetrically about $n=0$, and
energy
zero, so the Fermi level $n_F = (N-1)/2$.

Going back to the original wave functions, we have rederived the Weyl
character
formula (for $U(N)$ actually due to Schur):
\eqn\weyl{\chi_{\vec n}(\vec z) =
{\det_{1\le i,j\le N} z_i^{n_j} \over \det_{1\le i,j\le N}
z_i^{j-1-n_F}}.}
In terms of roots and weights, the indices $n_i$ with $n_1>n_2>\ldots>n_N$
are the components of the highest weight vector shifted by half the sum of
the
positive roots (usually denoted $\mu+\rho$) where the basis of the Cartan
subalgebra is just $(H_i)_{jk} = \delta_{ij}\delta_{jk}$.
In the language of Young tableaux, if $h_i$ is the number of boxes in the
$i$'th row, $n_i = (N-1)/2 + 1 - i + h_i$.

The $U(1)$ charge is $Q = \sum_i n_i$.  We can change this by a multiple
of $N$
by shifting all the fermions $n_i\rightarrow n_i+a$, but $Q\bmod N$ is
correlated with the conjugacy class of the $SU(N)$ representation (in
other
words the action of the center) reflecting the identification
$U(N) \cong SU(N)\times U(1)/\IZ_N$.

Interesting observables in this quantum mechanics, invariant under the
adjoint
action, are the invariant ``position'' operators
\eqn\wbasis{W_n = \tr U^n = \sum_i z_i^n}
and ``generalized Hamiltonians''
\eqn\ebasis{H_m = (-i)^m \sum_i {\partial^m\over\partial \theta_i^m}.}
For $m>2$ these are {\it not} the higher Casimirs $\tr E^m$ but are
polynomial
in them (see \cite[p. 163]{Zelobenko} for an explicit expression).  We
will not
discuss $m>2$ further here.

We next go to a second quantized formalism
with operators $B^{+}_{-n}$ and $B_n$ creating and destroying the fermion
mode
$z^n$, and $\psi(\theta)=\sum_n e^{in\theta} B_n$.  Then
$H = \int d\theta \partial\psi^{+} \partial\psi - E_0$.
The operators $W_n$ and $H_m$ will become fermion bilinears.

The first simplification of the large $N$ limit now appears.
If we never consider operators $W_n$ with $n\sim N$,
then fermions near the positive and negative Fermi surfaces completely
decouple.
We can then speak of a quasi-relativistic fermi system, with complex
chiral
left- and right-moving fermions.  We should also speak of $U$ raising the
left-moving (upper) fermions while lowering the right-movers, and $U^{-1}$
doing the opposite.
This suggests that we  refer to representations contained in tensor
products of
$O(N^0)$ fundamentals as ``chiral,'' and their complex conjugates as
``anti-chiral.''
The full representation theory is a product of chiral and anti-chiral
sectors.
So, let
$b^{+}_{n} = B^{+}_{-n_F-\epsilon+n}$,
$b_{n} = B_{n_F+\epsilon+n}$,
$\bar b^{+}_{n} = B^{+}_{n_F+\epsilon-n}$,
$\bar b_{n} = B_{-n_F-\epsilon-n}$,
where $\epsilon=\half$ is an choice of definition, introduced to give
antiperiodic ($n\in\IZ+\half$) moding for all $N$.
The local operators $\psi(z)=\sum_{n\in\IZ+\half} z^{-n} b_n$,
$\psi^{+}(z)$,
$\bar\psi(\bar z)=\sum_{n\in\IZ+\half} \bar z^{-n} \bar b_n$, and
$\bar\psi^{+}(\bar z)$
now satisfy standard 2d field theory commutation relations.
The standard Fock vacuum ($b_n|0>= b^{+}_n|0>=0$ for $n>0$) corresponds to
the
identity representation, and higher representations can be built by acting
with
the bilinears $b^{+}_{-n} b_{-m}$ and
$\bar b^{+}_{-n} \bar b_{-m}$.
Of these, clearly the simplest are the $W_n$'s which become
\footnote{Our CFT conventions are generally as in \cite{Itzyk}, except
that our
boson $\phi$ is $\sqrt{2}$ times theirs; in other words
$S_{\rm free} = \int d^2x (\partial\phi)^2/2\pi$.
Our contour integrals always have an implicit $1/2\pi i$,
and $\bar z$ integrals go counterclockwise.}
\eqal\wrel{W_n &=& \tr U^n \\
&=&\cint dz~ z^{-1-n} \psi^{+}(z) \psi(z) +
\cint d\bar z~ \bar z^{-1+n} \bar\psi^{+}(z) \bar\psi(z)\\
&=& \sum_m b^{+}_{n-m} b_{m} + \bar b^{+}_{m-n} \bar b_{-m}.
\label{wrel}}
We recognize the operators here as the left- and right-moving conformal
field
theory $U(1)$ currents, and the construction of the $W_n$'s as
bosonization:
\eqn\walph{
W_n \equiv \alpha_{-n} + \bar\alpha_{n}
= \int d\theta e^{in\theta}
\partial_\tau\phi(z=e^{n(\tau+i\theta)},\bar z=e^{n(\tau-i\theta)})}
defining the standard free boson oscillator expansion with
\eqal\bosalg{
\partial_z\phi(z)&=&i\sum_{m\in\IZ} \alpha_m z^{m-1}\no\\
{}[\alpha_m,\alpha_n]&=&[\bar\alpha_m,\bar\alpha_n]~=~m\delta_{m+n,0}\\
\left[\alpha_m,\bar\alpha_n\right]&=&0.\no}
Notice that the $W_n$ commute as operators, as they should.

The charges $\alpha_0$ and $\bar\alpha_0$ count fermion numbers.
Their sum is constant in our application.  There is a normal ordering
ambiguity
in the definition (\ref{wrel}) which we use to define it to be zero.
As for the difference $\alpha_0-\bar\alpha_0$, clearly it can be changed
by
operators like
\eqn\tn{\sum_n B^{+}_{-n_F-n+1} B_{-n_F+n} =
\sum_{n\in\IZ+\half} b^{+}_{-n} \bar b_{-n} =
\cint {dz\over z} \psi^{+}(z) \bar\psi(z^{*}).}
In the bosonic language it is winding number;
\eqn\wind{w = \alpha_0-\bar\alpha_0 = -i\cint dz \partial\phi
+i\cint d\bar z\bar\partial\phi
= -{1\over 2\pi}(\phi(2\pi)-\phi(0)).}

A better way to change the winding number is to turn it on continuously
from
zero; taking this back to the fermi picture we are continuously changing
the
fermion boundary conditions, or equivalently multiplying the wave function
by
\eqn\tmult{\psi(\vec z) \rightarrow
\left(\prod_{i=1}^N z_i\right)^s \psi(\vec z).}
Taking $s$ from zero to one gives a new state with the same $SU(N)$
quantum
numbers but $U(1)$ charge increased by $N$.
\def\boset{
The bosonization of the operator $T$ can be found using the standard
Mandelstam
representation for the fermions.
We make a left and right moving decomposition of the boson $\phi(z,\bar z)
=
\phi_L(z) + \phi_R(\bar z)$, then
$:e^{i\phi_L(z)}: = \psi(z)$, $:e^{-i\phi_L(z)}: = \psi^{+}(z)$, etc...
and
\eqn\boset{T = \cint {dz\over z} :\exp i(\phi_R(z)-\phi_L(z)):.}
}
We conclude that the original quantum mechanics on $U(N)$ is equivalent to
a
free bosonic field theory whose zero modes are treated rather
asymmetrically:
we sum over integer winding numbers, but not over momenta.

We should keep in mind that although the formalism so far suggests a close
relationship with two-dimensional conformal field theory, there is no a
priori
guarantee that the Hamiltonian or observables in a specific problem will
be
local in the two dimensions.  Of course the positions $\theta_i$
parameterize
the maximal torus of our original group, so local time evolution on the
group
manifold will reduce to local two dimensional evolution.  On the other
hand
group multiplication is an example of a natural operation with no locality
properties.  So not all problems will have simple conformal field theory
translations.

We could summarize by saying that the bose-fermi correspondance is the
large
$N$ limit of the Frobenius relation between characters and symmetric
polynomials.
A character $\chi_{\vec n}$ corresponds to a fermion Fock basis state in a
simple way; if one takes a ``chiral'' representation $R$,
with Young tableau with $h_i$ boxes in the $i$'th row, $1\le i\le r$,
this corresponds to the state
\eqn\state{|\vec h> = \chi_{\vec h}(U)|0> =
b^{+}_{\epsilon-h_1} b^{+}_{\epsilon-h_2} \ldots b^{+}_{\epsilon-h_r}
b_{-\epsilon} b_{-\epsilon-1} \ldots b_{-\epsilon-r+1} |0>.}
An alternate basis for class functions is
\eqn\class{\prod_i (\tr U^i)^{\sigma_i}}
(in terms of the $z_i$ these functions give a basis for the symmetric
polynomials)
which will correspond to a state built with bosonic operators
\eqn\clstate{
\label{clstate}
|\sigma> = \prod_i W_i^{\sigma_i} |0>.}
Orthonormality of the characters gives us
\eqn\res{\chi_{\vec n}(U)|0> = |\vec n> = \sum_\sigma <\vec n|\sigma>
W_i^{\sigma_i} |0>.}

Another application of this is the integration of class functions, which
is
just expectation values of products of operators.
For example,
\eqal\intex{\int dU (Tr U^{-2})(Tr U)^2  &=&
<0|(\alpha_2+\bar\alpha_{-2})(\alpha_{-1}+\bar\alpha_{1})^2|0>\\
&=& 0\no}
to all orders in $1/N$.
(and for finite $N>2$, by going back to the non-relativistic fermions.)

\def\finiteN{
Since the Frobenius relation exists for any finite $N$, we might wonder
whether
some form of bosonization exists at finite $N$.  The quantities $W_n$
satisfy
polynomial relations at finite $N$, }

The above was all for $U(N)$; the $U(1)$ generator is
$Q=H_1=\int d\theta \psi^{+}\partial\psi =\sum_n n B^{+}_{-n} B_n$ which
in the
large $N$ limit becomes
\eqal\uone{\label{uone}
Q&=& \sum_{n} n b^{+}_{-n} b_{n} - \sum_{n}n \bar b^{+}_{-n} \bar
b_{n}\no\\
&=& L_0 - \bar L_0}
(but see below.)
Constraining this to zero (keeping integer winding numbers) gives
representations of the quotient $SU(N)/\IZ_N$.
If we are interested in $SU(N)$ we have two options.
We can choose a representation of
$U(N)\cong SU(N)\times U(1)/\IZ_N$ for each representation of $SU(N)$, and
subtract the $U(1)$ part of
the second Casimir, $H_{U(1)}=Q^2/N$ from the Hamiltonian.
Or, we can extend our sum over states to get $SU(N)\times U(1)$.
A $U(1)$ character of charge $1$ is $\chi_1=\prod_i z_i^{1/N}$, so the
appropriate modification is simply to sum over winding numbers $k/N$, or
equivalently fermion sectors with twisted boundary conditions
$\psi(e^{2\pi
i}z) = e^{2\pi i(1/2+k/N)} \psi(z)$ and $\bar\psi(e^{-2\pi i}\bar z) =
e^{2\pi
i(1/2+k/N)} \bar\psi(\bar z)$.

Our quantum mechanical Hamiltonian is the second Casimir, which is not the
relativistic Hamiltonian $L_0+\bar L_0$.
In terms of relativistic fermions it is
\eqal\rham{H_{U(N)} &=& \sum_{n\in\IZ+\half} (n_F+\epsilon-n)^2
(b^{+}_{-n}b_n + \bar b^{+}_{n}\bar b_{-n}) - E_0\no\\
&=& NL_0 + N \bar L_0 + \sum_{n\in\IZ+\half} n^2
:b^{+}_{-n}b_n + \bar b^{+}_{-n}\bar b_{n}:\no\\
&=& NL_0 + N \bar L_0 + \cint dz~z^2 :\partial\psi^{+}\partial\psi:
+ \cint d\bar z~\bar z^2 :\bar \partial\bar \psi^{+}\bar \partial\bar
\psi:
\label{rham}}
(we know that the vacuum energy in this ground state is zero).

The bosonization of this Hamiltonian is very well known in the context of
matrix models, as it is just the Das-Jevicki-Sakita Hamiltonian
governing the dynamics of the eigenvalue density in hermitian matrix
quantum
mechanics.  We are retracing the steps of Gross and Klebanov and of Wadia
and
Sengupta \cite{GKWS}
to arrive at it.
The difference here is that there is no potential, and the fermions live
on a
circle.  We have a left and right moving decomposition of the boson
$\phi(z,\bar z) = \phi_L(z) + \phi_R(\bar z)$, and the standard formulas
$:e^{i\phi_L(z)}: = \psi(z)$, $:e^{-i\phi_L(z)}: = \psi^{+}(z)$, etc...
Substituting into (\ref{rham}), we can define the second derivative term
by
point-splitting the two operators and taking the limit.  The result must
be a
sum of operators of charge zero and dimension $(3,0)$ and $(0,3)$.  In
fact the
operators $\partial^3\phi$ or $\partial\phi\partial^2\phi$ would be
unimportant
here because they are total derivatives (and there are enough derivatives
to
kill the winding mode), so the only possibility is (the coefficients are
easily
checked on low lying states)
\eqal\jevsak{
\label{JS}
H &=& -{N\over 2}\cint dz~z:(\partial\phi)^2:
- {N\over 2}\cint d\bar z~\bar z:(\bar \partial\phi)^2: \\
&&+ {i\over 3}\cint d z~z^2 :(\partial\phi)^3:
+ {i\over 3}\cint d\bar z~\bar z^2 :(\bar\partial\phi)^3: \no\\
&\equiv& N L_0 + N \bar L_0 + H_I.\no}

The cubic interaction term in this Hamiltonian is quite natural, as we
could
see by considering the action of our original (\ref{qmham}) on states
(\ref{clstate}) --
it would contain terms preserving the ``string number'' (number of
traces), as
well as terms joining or splitting strings in higher order in $1/N$.
\footnote{See \cite{MinPoly} for a complete elaboration of this.}
$H_I$ is conserved under free time evolution (as are all the $H_m$'s).

Actually there is a slight awkwardness in the bosonic formalism at this
point:
with our present definitions, (\ref{uone}) is not quite correct.
The contribution of $w$ to the $U(1)$ charge is $N w$ and is correctly
reproduced by (\ref{uone}) only if we take the momentum $p=N$.
Although this might sound like a more natural choice, it obscures the
large $N$
limit: we will constantly need to expand $\partial\phi = N/2z + O(1)$ to
calculate.  Rather we take instead $p=0$ and
\eqn\uone{\label{uonemod}
Q= N w + L_0 - \bar L_0.}
This point is important only if we are interested in the operator $Q$;  in
particular (\ref{JS}) is correct with $p=0$.

All this could be done for a general group manifold.  Computing singlet
wave
functions again leads to the Weyl character formula.  For the groups
$Sp(2N)$
the maximal torus can be taken to be diagonal matrices
$\rm{diag}(z_i,z_i^{-1})$ and the Weyl group includes both permutations
and the
reflections $z_i\rightarrow z_i^{-1}$.  Although we will not try to
develop it
here, in the large $N$ limit this should produce a free fermion theory on
a
surface with boundary.
For $SO(N)$ at finite $N$ we would need an additional global degree of
freedom
to incorporate the spinor representations; however these have $C_2 \sim
N^2$ so
would drop out of our large $N$ considerations.

\newsec{YM$_2$ on the cylinder and torus}

This is really the same quantum mechanics on a group manifold under a
different
name.  Let us do canonical quantization with our space being a circle of
radius
$1$; time evolution will generate a cylinder of area $A=2\pi t$.
The Hamiltonian is $g^2 \int dx \tr E^2$, with
$E(x)^a = -i\partial/\partial A^a(x)$,
and we must impose Gauss' law $D_x E = 0$, which is solved by gauge
invariant
wave functions, i.e. satisfying
$\psi[g^{-1}(x)(\partial_x+A(x))g(x)] = \psi[A(x)]$.
A wave function is determined by its value on configurations of constant
$A(x)$, and gauge orbits are in one-to-one correspondance with values of
the
holonomy $U = P \exp i\int_0^{2\pi} A(x) dx$ modulo the adjoint action
$U\rightarrow g^{-1} U g$, completing the reduction to the singlet sector
of
quantum mechanics.
The standard large $N$ limit
is taken with gauge coupling $g^2 \sim 1/N$ and in two dimensions we can
set
$g^2 = 1/N$, defining our unit of length.
Then time evolution is generated by an $O(N^0)$ free Hamiltonian with
an $O(1/N)$ interaction term.
\cite{Wadia}
The ground state energy $E_0$ is freely adjustable, say by adding $E_0
\int
d^2x \sqrt{g}$ to our original Lagrangian.

The simplest physical quantity is the partition function on the torus,
\eqn\com{\Tr e^{-{2\pi t} H} = Z_{1\rightarrow 1} + O(N^{-2}).}
The leading term is $O(N^0)$ and the notation ``$1\rightarrow 1$''
indicates
that the string interpretation of this
\cite{Gross}
is a sum of (disconnected) maps from genus one world-sheets (at $N^0$) to
a
genus one target space.

Since the gauge invariant states correspond directly to our conformal
field
theory states, and the interaction is subleading,
$Z_{1\rightarrow 1}$ is almost the standard torus partition function of
free
$c=1$ conformal field theory.  The ``almost'' is there because the total
charge
of our fermi theory or momentum zero mode $\alpha_0+\bar\alpha_0$ of our
bose
theory is conserved; thus we have a partition function with this
constraint.
This is particularly easy to implement in the bose description; clearly
\eqn\etaeq{Z_{1\rightarrow 1} =  q^{E_0} \prod_{n\ge 1} {1\over
(1-q^{n})^2}
\sum_{w\in\IZ} q^{w^2}.}
where $q\equiv e^{-2\pi t}$.
The sum over winding modes is a rather uninteresting side effect of the
$U(1)$
factor.  The nicest way to eliminate it is to decouple the $SU(N)$ and
$U(1)$
in the way described above, by summing over winding numbers $w=k/N$.
In the large $N$ limit we clearly want to interpret such a sum as an
integral;
it is Gaussian, giving
\eqn\metaeq{\label{metaeq}
Z'_{1\rightarrow 1} = {N^2\over\sqrt{2\pi t}} q^{E_0} \prod_{n\ge 1}
{1\over
(1-q^{n})^2}.}
Another way of saying this is, we have an $SU(N)\times U(1)$ gauge theory
with
the same coupling constant in both sectors.  Using the coupling constant
which
gives a nice large $N$ limit, $g^2/N$, gives the extreme weak coupling
limit in
the $U(1)$ sector.  In this limit we cannot see the compact nature of the
group
$U(1)$.

Subleading corrections to this will have a string interpretation in terms
of
maps from higher genus world-sheets.  We can write an all-orders
expression
quite explicitly from the free fermion formalism: for $U(N)$,
\eqal\torus{Z_{\rm all\rightarrow 1}
&=& \sum_R e^{-A C_2(R)/N}\\
&=& q^{E_0} \oint {dz\over z}
\left[ \prod_{m\ge 1} (1 + z q^{m-1/2 + (m-1/2)^2/N})
\prod_{n\ge 1} (1 + z^{-1} q^{n-1/2 - (n-1/2)^2/N}) \right]^2 .\no}
The contour integral is there to implement the constraint of zero total
charge.
 It complicates the interpretation so again it is useful to do a bosonic
calculation.  From both the CFT and string points of view, the free energy
is a
sum over connected diagrams.  Expanding $\exp -{A\over N} H_I$ we have the
series
\eqn\gfe{
F_1(A) = F_{1\rightarrow 1} + \sum_{g\ge 2} N^{2-2g} F_{g\rightarrow 1}}
with
\eqn\freeg{
\label{freeg}
F_{g\rightarrow 1}(A) = {1 \over (2g-2)!}
< ({iA\over 3}\cint dz~z^2~ :\partial\phi(z)^3: + {\rm\ c.c.})^{2g-2} >_c
}
This is a connected correlation function on the torus (an annulus with $z$
and
$qz$ identified).  The integrals are taken over contours of constant
$|z|$, and
since $H_I$ is conserved, we can take $|z|$ to be slightly different for
each
contour, avoiding any possible singularities.  The Green's function
(defined by
the original oscillator expansion) will be the usual one
\cite[p. 571]{Itzyk}
if we use the same prescription as in (\ref{metaeq}) of integrating the
boson
winding mode:
\eqal\green{
G(\nu_1,\nu_2) &=& < \partial_\nu\phi(\nu_1) \partial_\nu\phi(\nu_2)
>\no\\
&=& \partial_1^2 \log \theta_1(\nu_1-\nu_2|\tau) + {\pi\over
t}\label{green}\\
&=& -\CP(\nu_1-\nu_2|\tau) - {\pi^2\over 3}E_2(\tau)+ {\pi\over t}.\no}
$z=e^{2\pi i\nu}$, $\tau=it$ and the Weierstrass function and Eisenstein
series
are defined in \cite{formulae}.  The $<\bar\partial\phi\bar\partial\phi>$
propagator is the same with $\nu\rightarrow\bar\nu$. Also
\eqn\mixed{<\partial\phi\bar\partial\phi> = -{\pi\over t}.}
If we want a group other than $SU(N)\times U(1)$, the zero mode
contribution
$\pi/t$ will be modified.

The first correction will be at $1/N^2$ from two insertions of our
interaction
Hamiltonian.
In changing variables from $z$ to $\nu$ we should remember that the normal
ordering of (\ref{JS}) was defined with respect to the $z$ coordinate.
Taking this into account however gives a contribution proportional to the
momentum, in other words zero.  The sum of terms involving contractions of
a
pair of operators from the same appearance of $H_I$ (in \cite{Gross},
contributions which can be disconnected by cutting a ``tube'') vanish (for
${\rm Re}~ \tau=0$). So,
\eqn\genustwo{\label{genustwo}
F_{2\rightarrow 1} = {2A^2\over 3(2\pi)^6}\int_0^1 d\nu\ G(\nu,0)^3.
}
The contour integrals are in the appendix, giving
\eqal\fres{\label{fres}
F_{2\rightarrow 1} &=& {A^2\over 2^5\cdot 3^4\cdot5}
(10 E_2^3 - 6 E_2 E_4 - 4 E_6 ) + {A\over 2^4\cdot9}(E_4 - E_2^2)\\
&=& A^2(8 q^2 + 64 q^3 + \ldots) + A(2 q + 12q^2+\ldots).\no}
In the interpretation of \cite{Gross} this is the generating function
counting
maps without folds from a genus two surface to a torus.  These can have
two
branch points (the $A^2$ term) or a ``handle'' (the $O(A)$ term).
Since the $U(1)$ piece does not contribute at subleading orders in $1/N$,
this
is exactly the $SU(N)$ result.
One can also see this by expanding $\partial\phi = w/z + \ldots$ and
integrating out $w$, which produces the correction to $H_I$ appropriate
for
$SU(N)$.

As for $F_{1\rightarrow 1}$, the most striking thing about this answer is
how
close it is to being a modular form (here of weight $6$) in the variable
$\tau$.
The Eisenstein series $E_{k}$ for $k\ge 4$ are forms of weight $k$, and
while
$E_2$ is not a form it has a very simple anomaly in its transformation
law:
\eqal\modu{
E_k({-1/\tau}) &=& \tau^k E_k(\tau),\hskip 1in k\ge 4\\
E_2({-1/\tau}) &=& \tau^2 (E_2(\tau) + {12/ 2\pi i\tau}).}
There is no analog of this at finite $N$; it is a non-trivial consequence
of
the quasi-relativistic nature of the degrees of freedom in the large $N$
limit.
Indeed, in terms of the original (unrescaled) couplings this is the
transformation $g^2 A/N \rightarrow N/g^2 A$, very different from familiar
strong-weak coupling duality.
It is a general property of the torus answers; from (\ref{freeg}) we see
that
$F_{g\rightarrow 1}/A^{2(g-1)}$ will ``almost'' be a modular form of
weight
$6(g-1)$.

It is very tempting to look for a string theory interpretation of this.%
\footnote{The following points were developed in discussions with D. Gross
and
C. Vafa.}
It is a target space duality invariance, like the $R\rightarrow 1/R$ of
the
free compact boson CFT.  In fact if the world-sheet embedding was
described by
a free (complex) boson we would expect precisely this symmetry.%
\footnote{See \cite{Bershadsky} for a more precise version of this.}
Furthermore, if we found that the (world-sheet) genus $g$ free energy was
precisely a modular form with weight proportional to $g-1$, we could
define a
combined transformation on area and string coupling which left the total
free
energy invariant, just as was the case for the compactified $c=1$
fundamental
string.
\cite{GrossKlebanov}

We should ask whether by changing definitions or modifying YM$_2$ slightly
we
could get a truly modular covariant answer.
{}From (\ref{metaeq}), it seems most promising to consider the
$SU(N)\times
U(1)$ case, though we have no deep understanding of why this choice of
group
should be better than $SU(N)$, say.  Adding a sum over momenta to
complement
the sum over integer windings appropriate for $U(N)$ does not seem
promising,
because the Hamiltonian has a term $p^3$, which would be unbounded below.

We would then like to make two changes: first, extend $\tau$ to a complex
parameter; second, extend the contour integrals in (\ref{freeg}) to
integrals
$\int d^2\nu$.
Ways to accomplish the first have been proposed by several physicists.
Since the area controls $L_0+\bar L_0$, we need to combine it with a
parameter
which controls $L_0-\bar L_0$.
Now in $D=2$ there is a theta term for $U(1)$ but not for $SU(N)$.
For the $U(N)$ theory we could just add the $U(1)$ theta term, which would
weigh the contribution of charge $q$ in the sum over representations by
$e^{iq\theta}$.  Modding out by $\IZ_N$ correlates this with the conjugacy
class in $SU(N)$, which in the large $N$ limit is just $L_0-\bar L_0$.
However this does not work for $SU(N)\times U(1)$, where there is no
correlation.
Instead we could consider twisted boundary conditions for the gauge field
on
the original torus.  A twist by a group element $C$ identifies holonomy
$U$ at
time $t$ with holonomy $C U$ at time $0$; gauge invariance
$\psi(U)=\psi(gUg^{-1})$ requires that $C$ be in the center, so there are
$N$
possible twists
$\theta=0,1/N,\ldots$ which act as $\exp{2\pi i\theta(L_0-\bar L_0)}$.  In
the
large $N$ limit we can consider $\tau=it+\theta$ to be a continuous
variable.

The other modification seems necessary if we want a modular covariant
answer,
but so far we have no real justification for doing this from the YM$_2$
point
of view.  Naively we would say that since $H_I$ is conserved we have
$A\int d\nu H_I = \int d^2\nu H_I$ but since (\ref{fres}) is not modular
covariant there must be a subtlety.  It is that we ignored short distance
singularities.  Integrating a meromorphic function $\int d^2\nu f(\nu)$
will
not produce divergences but we might get a finite piece which fixes things
up.
In fact, for $f(\nu)$ with a singularity only at zero, we can write
\eqn\newint{{1\over 2i}\int d\nu f(\nu) \wedge d(\bar \nu-\nu) =
{\rm Im}~\tau\int_0^1 d\nu f(\nu) + \pi {\rm Res}~\nu f(\nu)|_{\nu=0}}
and with this term
\eqn\newfree{
F^{inv}_{2\rightarrow 1} = {A^2\over 2^5\cdot 3^4\cdot5}~ {\rm Re}~
(10 \hat E_2^3 - 6 \hat E_2 E_4 - 4 E_6 )}
where $\hat E_2(\tau) = E_2(\tau) - 3/\pi{\rm Im}~\tau$ is modular
covariant.
The diagrams mentioned just above (\ref{genustwo}) now vanish.
This is different from (\ref{fres}) at $O(A)$ and now there are also terms
at
$O(A^0)$ and $O(A^{-1})$ whose string interpretation is unclear.
Furthermore, while the ``chiral'' and ``antichiral'' parts are separately
modular covariant, the combination appearing here is a bit strange.

We could certainly imagine other modifications of (\ref{fres}) to get
modular
covariance, but this modification does generalize to all genus and seems
relatively natural.  A modular covariant answer is a prerequisite for
comparison with topological string theory.
There is a striking similarity between the form of the collective field
theory
(\ref{JS}) and the Kodaira-Spencer field theory developed in \cite{BCOV},
describing topological string theory on a Calabi-Yau target space, which
may be
an important clue to the continuum string interpretation. \cite{Vafa}

\newsec{YM$_2$ on other Riemann surfaces}

Before discussing the formalism let us review some known results for the
large
$N$ partition function on other topologies.  The qualitative structure is
quite
clear from the expression as a sum over representations,
\eqn\repsum{\exp F_{\rm all\rightarrow G} =
\sum_R (\dim R)^{2-2G} \exp{-{A\over N} C_2(R)}, }
and the leading order behavior $\dim R \sim N^{n+\bar n}$ for a
representation
made from $n$ and $\bar n$ (anti)fundamentals.

For $G>1$, at order $1/N^{2s}$ only the finite set of representations with
$n+\bar n \le s$ can contribute.  Although there are interesting things to
say
about a string theory which reproduces these answers, since no observable
exhibits a sum over an infinite number of degrees of freedom, there is no
evidence for any non-topological field theory description.  In the present
framework we will see that although formally we can build higher genus
surfaces
by sewing cylinders, the ``vertex'' which accomplishes this is highly
non-local.

For $G=0$, there is an $O(N^2)$ contribution to $F$.  This is in many ways
the
most interesting case, because the leading behavior of the free energy for
higher dimensional Yang-Mills theory is $O(N^2)$.  It is quite possible
that
the qualitative behavior of the two-dimensional problem is of direct
relevance
for higher dimensions, because any higher dimensional space (and certainly
a
lattice as well) will contain embedded two-spheres.

Computing an $O(N^2)$ free energy in large $N$ is quite different from the
problems we treated above, because we can hope to find a saddle point
which
dominates the path integral.  Of course this is usually the reason we
think
that a large $N$ limit will simplify a problem; however it brings with it
a
complication: there can be more than one saddle point, and a phase
transition
where their actions are equal.

In terms of our present formalism we would expect
to describe the saddle point in terms of an expectation value for the
boson
$\phi$, determined by minimizing the Jevicki-Sakita action (\ref{JS}).
Although this makes sense it is much easier to describe the saddle point
in
terms of the conjugate variables $n_i$, since these are time-independent.
A
new feature of the problem is that since these are discrete, there is an
upper
bound on their density, and an associated phase transition when the bound
is
saturated.%
\cite{DK}

To return to our Hamiltonian formalism, the problem in this section is to
find
the wave functions for the disk and the three-holed sphere.
We might expect to be able to represent them as simple conformal field
theory
states, as is done in string field theory.
Let us start with the disk, or its zero-area limit, in other words the
wave
function $\psi=\delta(U)$.
Clearly acting on this state
$\Tr U^n = N ~~\forall n\ne 0$, or
\def\zeroname {D_0}
\eqn\diskcon{
\label{diskcon}
(\alpha_n + \bar\alpha_{-n} - N)|\zeroname> = 0.}
These constraints are easily solved:
\eqn\diskzero{|\zeroname> =
\exp ~-\sum_{n\ge 1}{1\over n}(\alpha_{-n}\bar\alpha_{-n} - N \alpha_{-n}
- N \bar\alpha_{-n}) ~~|0>.}
This state can be used to calculate the dimension of a representation:
\eqn\dimr{\dim R_{\vec n} = <\zeroname|\vec n>.}
(try for example the characters $\half((\Tr U)^2+\Tr U^2)$.)

Another characterization of this state is through boundary conditions of
the
fields -- we put a boundary $\tau=0$ with
$\partial\phi/\partial\tau = N\delta(\theta)$.
This is equivalent to taking Neumann boundary conditions, and inserting
the
operator $:\exp iN\phi(0):$.
In the fermi language this is a rather singular state, which is defined by
the
boundary condition that all the fermions (eigenvalues) are at $z=1$.
In the non-relativistic formalism, we can produce it by taking the o.p.e.
of
$N$ fermions, resulting in the state
$\prod_{i=0}^{N-1} \partial^i\psi~~|0>.$
Taking the inner product of (for example) a character with this state will
reproduce the calculation of the dimension of a representation by taking
the
limit of all $z_i\rightarrow 1$ in the Weyl character formula using
l'H\^opital's rule.

The  YM${}_2$ sphere free energy is then
\eqal\free{ \exp F_{{\rm all}\rightarrow 0} &=& <D_0|e^{-A H}|D_0> \\
&=& <:e^{-iN\phi(\tau)}:
{}~e^{{t\over N}H_I}~ :e^{iN\phi(0)}:>_{\rm cyl}\no
}
a correlation function on a cylinder of length $\tau$.
We see that we cannot expand $\exp {t\over N}H_I$; the $1/N$ is
compensated in
correlation functions by the explicit $N$ in the vertex operators.  On the
other hand the problem is classical; we can rescale $\phi\rightarrow
N\phi$ and
get an overall $N^2$ in the exponential.

The qualitative behavior of the dominant classical solution is clear.  We
have
a fluid with a conserved particle number; the boundary condition is that
at
time $t=0$ it is concentrated at $\theta=0$; as it evolves it spreads out
but
must recontract from $t=\tau/2$ on to meet the boundary condition at
$t=\tau$.
This is not the easiest way to solve the problem, which is treated more
simply
in \cite{DK}, but it does make the nature of the large $N$ transition
clear:
at a critical $\tau_c$, the two edges of the particle distribution meet at
$t=\tau_c/2$ and $\theta=\pi$.  For $\tau\le\tau_c$ they do not know they
live
on a circle; beyond this value it is crucial, and the classical solution
is a
non-analytic function of $\tau$ at this point.
(See also \cite{MP+Caselle}.)

The expansion of \cite{Gross} reproduces the $\tau\ge\tau_c$ behavior but
has
no apparent relation with the $\tau < \tau_c$ behavior.  Putting back the
$g^2$
dependence shows that it is valid at strong coupling.
Now the main reason for the transition was simply the compactness of the
gauge
group, just as for the Gross-Witten transition \cite{GW}, which suggests
that
higher dimensional theories will have the same problem.
This is not just a property of an approximation but of the true continuum
result, and the dynamical picture above suggests the strange possibility
that
in higher dimensions we could have non-analyticity in the local
observables at
large $N$, for example in the dependence of a Wilson loop expectation
value on
its length.  (This is not the case in $D=2$, see \cite{Daul+Boulatov}).

To try to understand the effects of the various world-sheet features of
\cite{Gross}, we could ask how modifying them or removing them affects the
answers.  Comparison with the derivation there shows that the
$\Omega$-points
of \cite{Gross} are represented here by the states $|D_0>$
while the ``movable branch points'' (those which come with a weight
proportional to the area of the world-sheet) correspond to insertions of
the
interaction $H_I$ (as we saw above for the torus target space).  On the
sphere
we get an interesting result even after dropping the movable branch
points.
\footnote{This modification has also been considered by D. Gross.}
Since we could accomplish this by multiplying $H_I$ by a new coupling
constant
and taking it to zero, the resulting theory is in a sense continuously
connected with the standard YM$_2$.
In fact this modification corresponds to a local modification of the
original
YM$_2$ action.
\cite{Zhao}
Thus it is not inconceivable that the corresponding modification of a
$D>2$-dimensional string theory would be in the same universality class as
YM$_D$.
We might take this as a working hypothesis in interpreting string rules
like
those of \cite{Gross,Kostov}: features we cannot change by a local
modification
of the original Yang-Mills action (like the $\Omega$ points) are more
likely to
be relevant for physics than features (like the movable branch points)
that can
be so changed.

What makes it all the more interesting is that in $D=2$ this modification
pushes the transition to zero coupling!  For simplicity drop the winding
modes
of the boson, so this result exactly enumerates covers of the sphere
branched
only at the $\Omega$ points.
Correlation functions on the cylinder with Neumann boundary conditions are
equal to those of a chiral boson on the doubled surface, here a torus of
modulus $2\tau$.  Then
\eqal\free{ \exp F^{\rm modified}_{{\rm all}\rightarrow 0}&=&
 <:e^{-iN\phi(\tau)}: ~:e^{iN\phi(0)}:>_{\rm cyl}\no\\
&=& {1\over\eta(2\tau)} \exp N^2 \left[ <\phi(\tau)\phi(0)> -
\lim_{z\rightarrow 0} (<\phi(z)\phi(0)>+\log z) \right]\no\\
 &=& {1\over\eta(2\tau)} \exp -N^2 \log
{\theta_1(\tau|2\tau)\over\theta_1'(0|2\tau)} \\
 &=& \prod_{m\ge 1}{1\over(1-q^{2m})} \left({1-q^{2m}\over
1-q^{2m-1}}\right)^{2N^2}.\no }
This is an analytic function for all real $\tau>0$, with an expansion in
$\exp -\tau$ whose terms have a string representation, and whose
$\tau\rightarrow 0$ behavior is remarkably similar to the ``correct''
(heat
kernel) weak coupling behavior $F\sim -\half N^2 \log \tau$.
Whether this is of any relevance to $D>2$ is unclear, and the action which
defines this theory is quite peculiar, but this suggests that the heat
kernel
action might not be the last chapter in the story.

We briefly return to $G>1$.
We can build higher dimensional surfaces by sewing, and the appropriate
``pants'' vertex is well known in the character language:
\eqn\vert{<V_3|~|\psi_1>~|\psi_2>~|\psi_3> =
 \int dU dV dW \psi_1(UVU^{-1}W) \psi_2(V^{-1}) \psi_3(W^{-1})}
\eqn\verthree{|V_3> = \sum_R {1\over \dim R} |\chi_R>_1 |\chi_R>_2
|\chi_R>_3.
\label{verthree}}
Since this expression involves group multiplication, a Ward identity
analogous
to (\ref{diskcon}) would be highly non-local in our two-dimensional
auxiliary
space.  The result is clearly non-local at each order in $1/N$ as well; it
is a
sum over finitely many states which separately have no local definition.
All this and consideration of what sort of vertex could satisfy the
equation
$<D_0|V_3> = 1$ leads to the conclusion that there is not likely to be an
expression for the vertex much simpler than (\ref{verthree}).

A final comment, which we will not apply here
(but see \cite{Doug}).
Clearly there are many other group theoretic ideas we could try to fit
into
this formalism.  One interesting one is multiplication of characters,
which
allows computing tensor product decompositions.  Evidently it is much
easier to
multiply symmetric functions expressed in the basis of power sums
(\ref{clstate}), so this will be a simpler operation in the bosonic
language.
Since we are multiplying wave functions at the same point on the group
manifold, we expect this to be a local operation in our two dimensional
CFT
language.
In fact we can write multiplication of $n$ wave functions as a vertex
$|V_n>$,
determined by the condition that
\eqn\wchar{<\psi_1|<\psi_2|\ldots \Tr U^k_i |V_n> =
\int dU~ \Tr U^k \prod_i \psi_{i}(U)}
is the same no matter which wave function we multiply by $\Tr U^k$.
In conformal field theory terms this means that we have a boundary on
which $n$
cylinders meet, and the local boundary condition
\eqn\wver{\partial_\tau\phi_i(z)=\partial_\tau\phi_j(z) \qquad\forall
i,j.}
This condition only couples mode $n$ to $-n$ and is easy to translate
into an oscillator expression for the vertex
\eqn\vert{|V_n> = \exp \sum_{i\ne j}
\sum_{n\ge 1} {1\over n} \alpha_{-n}^{(i)} \bar\alpha_{-n}^{(j)} ~~|0>.}
In a certain sense this is a trivial result; however it would be quite
amusing
to translate it into the fermionic language, because it would amount to a
new
proof of the Littlewood-Richardson rule for tensor product decompositions,
at
least in the large $N$ limit.
\medskip

Some contour integrals used in section 3: \cite{formulae}
\eqal\contour{
&&\int_0^1 d\nu\ \partial^2_\nu \log\theta_1(\nu) = 0 \no\\
&&\int_0^1 d\nu\ \CP(\nu)^2 = {\pi^4\over 9} E_4 \no\\
&&\int_0^1 d\nu\ \CP(\nu)^3 = {4\pi^6\over 5\cdot 27} E_6
-{\pi^6\over 15} E_4 E_2 \no}

\medskip
I thank T. Banks, G. Moore, D. Gross, V. Kazakov, J. Minahan, J.
Polchinski, R.
Rudd, A. Strominger, and C. Vafa for enjoyable discussions, and DOE grant
DE-FG05-90ER40559, NSF grants PHY-9157016 and PHY89-04035, and the Sloan
Foundation for their support.

\index{QCD!2-dimensional}
\index{QCD$_2$}
\index{Conformal field Theory CFT}
\index{Duality!$R\rightarrow 1/R$}
\index{Gauge theory}
\index{Matrix models}
\index{Yang-Mills theory}
\index{Large $N$}


\begin{thebibliography}{99}
\bibitem{Jimbo} M. Jimbo and T. Miwa, in Integrable Systems in Statistical
Mechanics, eds. G. M. D'Ariano, A. Montorsi and M. G. Rasetti, 1985;
World Scientific, Singapore.
\bibitem{Segal} Loop Groups, A. Pressley and G. Segal, Oxford, 1986; most
relevant to the present discussion are chapters 10 and 14.3.\\
Bombay lectures on highest weight representations of infinite
dimensional Lie algebras, V. G. Kac and A. K. Raina,
World Scientific, 1987.
\bibitem{DMP+Jevicki+Stone} M. Stone, Phys. Rev. B42 (1990) 8399.\\
A. Jevicki, Nucl. Phys. B376 (1992) 75.\\
R. Dijkgraaf, G. Moore and R. Plesser,
Nucl. Phys. B394 (1993) 356.
\bibitem{Polchinski} A good review is by J. Polchinski,
presented at the Symposium on Black Holes, Wormholes, Membranes and
Superstrings,
Houston, TX, Jan 16-18, 1992, hep-th/9210045.
\bibitem{kkone} V. A. Kazakov and I. K. Kostov, Nucl. Phys. B176
(1980) 199.
\bibitem{Gross} D. J. Gross and W. Taylor, Nucl. Phys. B400 (1993) 181 and
 Nucl. Phys. B403 (1993) 395.
\bibitem{Kostov} I. K. Kostov, SACLAY-SPHT-93-050, June 1993.
hep-th/9306110.
\bibitem{Bershadsky} M. Bershadsky, S. Cecotti, H. Ooguri, and C. Vafa,
Nucl.
Phys. B405 (1993) 279.
\bibitem{DR} R. Dijkgraaf and R. Rudd, unpublished.
\bibitem{Ambjorn}
See J. Ambjorn et. al., Nucl. Phys. B270 (1986) 457 and references there.
\bibitem{BIPZ} E. Br\'ezin, C. Itzykson, G. Parisi and J.-B. Zuber,
Comm.~Math.~Phys. 59 (1978) 35.
\bibitem{HCandHT} Harish-Chandra, Amer. J. Math. 79 (1957) 87-120; for a
nice
pedagogical treatment which continues with non-compact groups see
Non-Abelian
Harmonic Analysis, R. Howe and E. C. Tan, Springer-Verlag, 1992.
\bibitem{Dowker} J. S. Dowker, J. Phys. A (1970) 451.
\bibitem{Zelobenko} Compact Lie Groups and their Representations, D. P.
Zelobenko, AMS translations vol. 40, AMS, 1973.
\bibitem{Itzyk} Statistical Field Theory, C. Itzykson and J.-M. Drouffe,
Cambridge, 1989.
\bibitem{GKWS} D. J. Gross and I. Klebanov, Nucl. Phys. B352 (1990) 671;\\
A. M. Sengupta and S. R. Wadia, Int. J. Mod. Phys. A6 (1991) 1961.
\bibitem{Wadia} S. R. Wadia, Phys. Lett. 93B (1980) 403.
\bibitem{MinPoly} J. A. Minahan and A. P. Polychronakos,
Phys. Lett. B312 (1993) 155.
\bibitem{formulae}
Introduction to Elliptic Curves and Modular Forms, N. Koblitz,
Springer-Verlag,
1984. \\
Table of Integrals, Series and Products, I. S. Gradshteyn and I. M.
Ryzhik,
Academic Press, 1980.
\bibitem{GrossKlebanov} D. J. Gross and I. Klebanov, Nucl. Phys. B344
(1990)
475.
\bibitem{BCOV} M. Bershadsky, S. Cecotti, H. Ooguri and C. Vafa,
HUTP-93-A025,
Sep. 1993.  hep-th/9309140.
\bibitem{Vafa} C. Vafa, private communication.
\bibitem{DK} M. R. Douglas and V. A. Kazakov, to appear in Phys. Lett. B;
see also Kazakov's lecture in this volume.
\bibitem{MP+Caselle} J. A. Minahan and A. P. Polychronakos,
CERN-TH-7016-93, Sept. 1993.  hep-th/9309119. \\
M. Caselle, A. D'Adda, L. Magnea, S. Panzeri, DFTT-50-93,
Sept. 1993. hep-th/9309107.
\bibitem{GW} D. J. Gross and E. Witten, Phys. Rev. D21 (1980) 446.
\bibitem{Daul+Boulatov} J.-M. Daul and V. A. Kazakov, LPTENS-93-37, Oct.
1993.
hep-th/9310165.\\
D. V. Boulatov, NBI-HE-93-57, Oct. 1993.  hep-th/9310041.
\bibitem{Zhao} For a discussion of such modifications, see W.-D. Zhao,
PUPT-1390, Mar. 1993, or upcoming work by M. R. Douglas, K. Li, and M.
Staudacher.
\bibitem{Doug} M. R. Douglas, RU-93-13, NSF-ITP-93-39, Mar. 1993.
hep-th/9303159.

\end{thebibliography}
\end{document}